\pdfoutput=1
\RequirePackage{ifpdf}
\ifpdf % We~are running pdfTeX in pdf mode
\documentclass[pdftex]{sigma}
\else
\documentclass{sigma}
\fi

\DeclareMathOperator{\sgn}{sgn}
\newcommand\abs[1]{\lvert #1 \rvert}

\numberwithin{equation}{section}

\newtheorem{Theorem}{Theorem}[section]
\newtheorem{Corollary}[Theorem]{Corollary}
\newtheorem{Lemma}[Theorem]{Lemma}
 { \theoremstyle{definition}
\newtheorem{Remark}[Theorem]{Remark} }

\usepackage{bbm}

\begin{document}
\allowdisplaybreaks

\newcommand{\arXivNumber}{2103.10495}

\renewcommand{\PaperNumber}{074}

\FirstPageHeading

\ShortArticleName{Non-Integrability of the Kepler and the Two-Body Problems on the Heisenberg Group}

\ArticleName{Non-Integrability of the Kepler and the Two-Body\\ Problems on the Heisenberg Group}

\Author{Tomasz STACHOWIAK~$^{\rm a}$ and Andrzej J.~MACIEJEWSKI~$^{\rm b}$}

\AuthorNameForHeading{T.~Stachowiak and A.J.~Maciejewski}

\Address{$^{\rm a)}$~Krak\'ow, Poland}
\EmailD{\href{mailto:tomasz@monodromy.group}{tomasz@monodromy.group}}
\URLaddressD{\url{https://monodromy.group}}

\Address{$^{\rm b)}$~Janusz Gil Institute of Astronomy, University of Zielona G\'ora,\\
\hphantom{$^{\rm b)}$}~Licealna 9, PL-65--417 Zielona G\'ora, Poland}
\EmailD{\href{mailto:a.maciejewski@ia.uz.zgora.pl}{a.maciejewski@ia.uz.zgora.pl}}

\ArticleDates{Received May 04, 2021, in final form July 27, 2021; Published online July 31, 2021}

\Abstract{The analog of the Kepler system defined on the Heisenberg group introduced by Montgomery and Shanbrom in [\textit{Fields Inst.\ Commun.}, Vol.~73, Springer, New York, 2015, 319--342, arXiv:1212.2713] is integrable on the zero level of the Hamiltonian. We show that in all other cases the system is not Liouville integrable due to the lack of additional meromorphic first integrals. We prove that the analog of the two-body problem on the Heisenberg group is not integrable in the Liouville sense.}


\Classification{37J30; 70F05; 70H07; 70G45; 53C17}

\section{Introduction}
The idea of studying the Kepler problem, or the problem of $n$ bodies, in a non-Euclidean space has a long history, but in almost all cases such
generalisations were performed in spaces of~con\-stant curvature. For a very detailed and critical recent overview
of this subject we refer to the nice article \cite{borisov:16::}.

In \cite{1} the authors considered the question of generalization of
the Kepler problem and its resulting mechanics, based on first principles. The idea was to
formulate an analog of the classical problem in spaces which are
homogeneous, isotropic and admit dilations. The last requirement is
very restrictive because among homogeneous Riemannian manifolds only
Euclidean ones admit dilations. The simplest non-Euclidean metric
space satisfying all the required properties is the Heisenberg group
of special upper-triangular matrices
\begin{equation*}
 \begin{pmatrix}
 1 & x_1 & x_3 \\ 0 & 1 & x_2 \\ 0 & 0 & 1
 \end{pmatrix}\!,
 \qquad x_i\in\mathbb{R}.
\end{equation*}
An isomorphic representation is obtained by taking new coordinates
\begin{equation*}
 x_1 = x, \qquad x_2 = y, \qquad x_3 = z +\frac12 x_1 x_2,
\end{equation*}
in which the group action is
\begin{equation}
 (x_1,y_1,z_1)\cdot(x_2,y_2,z_2)=\bigg(x_1+x_2,y_1+y_2,z_1+z_2+\frac12(x_1 y_2-x_2 y_1)\bigg).
 \label{groupop}
\end{equation}

This manifold carries a maximally non-integrable distribution spanned
by the vector fields
\begin{equation*}
 X_1 = \partial_x - \frac12 y\partial_z,\qquad
 X_2 = \partial_y + \frac12 x\partial_z,
\end{equation*}
and which define the sub-Riemannian structure. Furthermore, there
exists a sub-elliptic Laplacian $\Delta:=X_1^2+X_2^2$ for which the
fundamental solution to the Poisson equation $\Delta U=\varrho$, with density $\varrho$, is known~\cite{Folland}. This makes the analogy complete because one can
construct the kinetic energy with the sub-Riemannian metric, and take
the potential as the point-source solution $U$;
the Hamiltonian of such Kepler--Heisenberg system is then
\begin{equation}
 \label{KH}
 H = \frac12 \bigg(p_x - \frac12 y p_z\bigg)^2 + \frac12\bigg(p_y+\frac12 x p_z\bigg)^2
 - \frac{\kappa}{\rho}, \qquad
 \rho:=\sqrt{(x^2+y^2)^2+{16}z^2},
\end{equation}
where $\kappa$ is a non-zero real parameter. The properties of this
system have been analysed in \cite{dots, 1,2}, where it was shown that it
has an invariant submanifold given by $z=0$, $p_z=0$,
$p_{\theta} = x p_y -y p_x=0$ on which all trajectories are straight
lines. More importantly, it was also demonstrated that all periodic
solutions must lie on the zero-energy level and that the whole system
is Liouville integrable there. This happens because of the quantity
$J=x p_x+ y p_y +2 z p_z$, for which $\dot{J} = 2H$, and on the level
$H=0$, it is the third integral of motion next to $H$ and
$p_{\theta}$, with which it commutes.
\begin{Remark}\label{rem1}
 The form of potential in~\eqref{KH} is the same as in~\cite{dots} which is the correction of that~in~\cite{1}.
\end{Remark}
We complete the above findings by proving that for all the other
values of energy, the system is not integrable. In addition, the generalisation of the above system to two bodies is straightforward, and we prove its non-integrability as well.

\section{The Kepler problem}

In this section we prove that the Kepler problem on the Heisenberg group is not integrable.
This result will follow as a corollary from a more general theorem.
Our proof is based on the Morales--Ramis theorem \cite{Morales:01::b1} which gives
the following necessary conditions for the integrability.
\begin{Theorem} \label{thm:MoRa}
 If a Hamiltonian system is Liouville integrable, with meromorphic
 first integrals, then the identity component of the differential
 Galois group of variational equations along any non-constant
 solution is Abelian.
\end{Theorem}

Two remarks are in order before we jump to application. One is the
``fine print'' of the above theorem: if the variational equation is
not Fuchsian, then only rational first integrals can be~treated. This
will turn out to be the case here, due to the choice of the particular
solution.

Secondly, because the Hamiltonian~\eqref{KH} itself is not
meromorphic, thanks to algebraic potential
\begin{equation*} %\label{potkep}
 V= - \frac{\kappa}{\sqrt{\big(x^2+y^2\big)^2+{16}z^2}},
\end{equation*}
a slight modification to Theorem~\ref{thm:MoRa} is necessary. We describe it in Appendix~\ref{App0}.

When applying the above, the main difficulty is connected
with determination of properties of the differential Galois group of
variational equations. If they can be reduced to a second order equation, then we can use the decisive Kovacic
algorithm~\cite{3}. Sometimes it is possible to~show that the
considered variational equations contain as a subsystem an equation
for which the differential Galois group is known, e.g., hypergeometric
equation and its confluent form. Here~we give a very useful example
which we will apply later.
\begin{Theorem}[H.P.~Rehm, 1979] \label{thm:par}
 Assume that complex parameters $\alpha\neq 0$, $\beta$, and $\gamma$
 of the parabolic cylinder equation
 \begin{equation} \label{eq:5}
 w''(z) - \big(\alpha^2 z^2 +2 \alpha \beta z + \gamma\big) w(z)=0
 \end{equation}
 are such that $\big(\beta^2-\gamma\big)/\alpha$ is not an odd integer. Then
 its differential Galois group is $\mathrm{SL}(2,\mathbb{C})$.
\end{Theorem}
This theorem was proved in \cite{Rehm} and later in~\cite{baby} it was
proved in another way with the help of the Kovacic algorithm.

Now we are ready to formulate the following theorem.
\begin{Theorem} %\label{thm:gen}
 Let us consider the system given by the Hamiltonian function
 \begin{equation}
 \label{eq:gen}
 H = \frac12 \bigg(p_x - \frac12 y p_z\bigg)^2 + \frac12\bigg(p_y+\frac12 x p_z\bigg)^2
 + V(x,y,z),
 \end{equation}
 where $V(x,y,z)=W(z,\rho)$, and $W(z,\rho)\in\mathbb{C}(z,\rho)$ is a
 rational function with
 \[
 \rho=\sqrt{\big(x^2+y^2\big)^2+{16}z^2}.
 \]
 If there exits a nonzero $c\in \mathbb{C}$ such
 that
 \begin{equation}
 \label{eq:con}
2a:= \frac{\partial V}{\partial z}(0,0,c) \neq 0,
 \end{equation}
 then the system is not integrable in the Liouville sense with first
 integrals which are rational functions of
 $(x,y,z, \rho, p_x,p_y,p_z)$.
\end{Theorem}
\begin{proof}
 The Hamilton's equations generated by~\eqref{eq:gen} have the
 particular solution
 \begin{equation*}
 \varphi(t) = [x(t),p_x(t),y(t),p_y(t), z(t),p_z(t)] =
 [0, 0,0,0, c, -2 a t],
 \end{equation*}
 where $a\neq 0$ is defined by~\eqref{eq:con}. The system linearized
 along this solution reads
 \begin{equation} \label{VE}
 \dot\eta = \begin{bmatrix}
 0 & 1 & a t & 0 & 0 & 0 \\
 -a^2 t^2 & 0 & 0 & at & 0 & 0 \\
 -a t& 0 & 0 & 1 & 0 & 0\\
 0 & -a t & - a^2 t^2 & 0 & 0 & 0\\
 0 & 0 & 0 & 0 & 0 & 0\\
 0 & 0 & 0 & 0 & C & 0 \end{bmatrix} \eta,
 \end{equation}
 where the variations of $[x,p_x,y,p_y,z,p_z]$ are
 $[\eta_1,\eta_2,\eta_3,\eta_4,\eta_5,\eta_6]$, respectively, and
 explicit form of $C$ is irrelevant for further considerations.

 Note that $\varphi(t)$ is constant in the configuration space but
 not in the phase space, and that the solvable subsystem for $\eta_5$
 and $\eta_6$ separates completely. It is sufficient to consider only
 the remaining components which form the so-called normal variational
 equations. It is easy to show that if the system is integrable then
 the identity component of the normal variational equations is
 Abelian. Now the problem is that generally for four-dimensional systems (or
 equations of order four) of arbitrary origin there is no decisive algorithm which
 allows to determine their differential Galois group or the identity component of this group. Sometimes a
 higher-dimensional system splits into systems of lower dimensions or
 contains as a subsystem of lower dimension. In such case we say that
 the system is reducible and the problem is reduced to a simpler one.

 Some simplification can be achieved by reverse-engineering the solution described below in~Remark~\ref{rem1}, but there is a more systematic approach: to check for factorisation. This can be done algorithmically, and we include the general outline in Appendix~\ref{App1}. In order to obtain a~particularly simple splitting of the variational equation, we modified the resulting transformation slightly, and performed the
 following non-canonical change of variables
 \begin{alignat}{3}
& q_1 = x +\mathrm{i}y, \qquad &&h_1=p_x+\mathrm{i}p_y + \frac{\mathrm{i}}{2}p_{z}( x +\mathrm{i}y),& \nonumber
 \\
& q_2 = x -\mathrm{i}y, \qquad &&h_2=p_x-\mathrm{i}p_y - \frac{\mathrm{i}}{2}p_{z}( x -\mathrm{i}y),&\nonumber
 \\
& q_3= z, \qquad &&h_3 = p_z.&\label{eq:1}
 \end{alignat}
 In new variables, the equations of motion read
 \begin{alignat*}{3}
 %\label{eq:2}
 &\dot q_1 = h_1, \qquad&&
 \dot h_1 = \mathrm{i} h_1h_3
 -\frac{\mathrm{i}}{2}q_1\bigg[ \frac{\partial W}{\partial
 z}(q_3, \chi) + \frac{4(4q_3 -\mathrm{i}q_1q_2 )}{\chi}
 \frac{\partial W}{\partial \rho}(q_3, \chi) \bigg],&
 \\
 &\dot q_2 = h_2, \qquad&&
 \dot h_2 = -\mathrm{i} h_2 h_3 +
 \frac{\mathrm{i}}{2}q_2\bigg[ \frac{\partial W}{\partial z}(q_3,
 \chi) + \frac{4(4q_3 +\mathrm{i}q_1q_2) }{\chi}
 \frac{\partial W}{\partial \rho}(q_3, \chi) \bigg],&
 \\
 &\dot q_3 =\frac{\mathrm{i}}{4}(q_1h_2-q_2h_1), \qquad&&
 \dot h_3=
 -\frac{\partial W}{\partial z}(q_3, \chi) - 16\frac{q_3}{\chi}
 \frac{\partial W}{\partial \rho}(q_3, \chi) ,&
 \end{alignat*}
 where $\chi=\sqrt{q_1^2q_2^2 +16q_3^{2}}$. The
 considered particular solution of these equations is
 \begin{equation*}
 \varphi(t) = \big[q_1(t),h_1(t),q_2(t),h_2(t), q_3(t),h_3(t)\big] =
 [0, 0,0,0, c, -2 a t ],
 \end{equation*}
 where $c\neq 0$, and the existence of a non-zero $a$ is guaranteed by~\eqref{eq:con}. Now the
 variational equations have the form
 \begin{equation}
 \label{eq:3}
 \dot \eta = A \eta, \qquad
 A:=\begin{bmatrix}
 A_1 & 0 & 0 \\
 0 & A_2 & 0\\
 0 & 0 & A_3
 \end{bmatrix}\!,
 \end{equation}
 where
 \begin{equation*} %\label{eq:4}
 A_1:=\begin{bmatrix}
 0 & 1 \\
 -\mathrm{i}a & -2 \mathrm{i}a t
 \end{bmatrix}\! , \qquad
 A_2 = A_1^{\star}, \qquad
 A_3:=\begin{bmatrix}
 0 & 0 \\
 C & 0
 \end{bmatrix}\! .
 \end{equation*}

 The subsystem corresponding to variables $(\eta_1,\eta_2)$ reads
 \begin{equation*}
 %\label{eq:7}
 \dot\eta_1=\eta_2, \qquad \dot\eta_{2 }= -\mathrm{i}a\eta_1 -2 \mathrm{i}a t \eta_{2}.
 \end{equation*}
 We rewrite it as a second order equation
 \begin{equation}
 \label{eq:8}
 \eta_1'' +2 \mathrm{i}a t \eta_1' + \mathrm{i}a\eta_1=0.
 \end{equation}
 Making the following change of dependent variable
 \begin{equation}
 \label{eq:9}
 \eta_1(t) = w(t)\mathrm{e}^{-\mathrm{i}a t^2/2}
 \end{equation}
 we obtain the reduced form of equation~\eqref{eq:8}
 \begin{equation}
 \label{eq:10}
 w''(t) + a^2 t^2 w(t)=0.
 \end{equation}
 The transformation~\eqref{eq:9} does not change the identity
 component of the differential Galois group of the equation. Now,
 equation~\eqref{eq:10} is a particular case of parabolic cylinder
 equation~\eqref{eq:5} with $\alpha^2 = -a^2$ and $\beta=\gamma=0$,
 so, by Theorem~\ref{thm:par}, its differential Galois group is
 $\mathrm{SL}(2,\mathbb{C})$. As~$\mathrm{SL}(2,\mathbb{C})$ is
 connected its identity component is the whole group. So, it is not
 Abelian, and, by~Theorem~\ref{thm:MoRa}, the system is not
 integrable.
\end{proof}
Condition~\eqref{eq:con} expressed by function $W(z,\rho)$ reads
 \begin{equation}
 \label{eq:conW}
2a:= \frac{\partial W}{\partial z}(c,4\abs{c}) +
 {4}\frac{c}{\abs{c}}\frac{\partial W}{\partial \rho}(c,4\abs{c})\neq 0.
 \end{equation}
Thus if $W(z,\rho)$ depends only on $\rho$, that is $W$ is a
non-constant rational function of $\rho$, then
condition~\eqref{eq:conW} is satisfied. Since for the Kepler potential
$W(z,\rho)=-\kappa/\rho$, the above theorem proves in particular:
\begin{Corollary}
The Kepler--Heisenberg problem, as formulated by Montgomery and Shanbrom in~{\rm \cite{1}}, is not integrable in the Liouville sense with rational first integrals.
\end{Corollary}

\begin{Remark} %\label{rem:bessl}
 It is worth noticing that the general solution of
 equation~\eqref{eq:8} is
 \begin{equation*}
 %\label{eq:6}
 \eta_1(t)= \sqrt{z} \mathrm{e}^{-\mathrm{i}a t^2/2}
 \Big[ C_1 J_{\frac{1}{4}} \big(a t^2\big) + C_2 Y_{\frac{1}{4}} \big(a t^2\big) \Big],
 \end{equation*}
 where $J_{\alpha}(z)$ and $Y_{\alpha}(z)$ are Bessel functions of
 the first and second type, respectively; $C_1$ and $C_2$ are
 arbitrary complex constants. At this point it becomes clear that the
 Galois group cannot be solvable, because the Bessel functions are
 Liouvillian only when their order is half an odd integer \cite{3}.
\end{Remark}

\section{Two-body problem}

Having dealt with the original generalisation of the Kepler problem
proposed in \cite{1}, a natural question arises about the two-body
problem. In the classical Kepler problem, there is no fundamental
difference between one and two bodies: the latter still leads to the
Kepler problem for a single body of reduced mass, revolving around the
center of mass. The reduction is possible due to the symmetries of the Euclidean space, which generate boosts, and correspond closely to the motion: relative positions of two particles follow a geodesic. This is not the case for the Heisenberg group, where the group operation \eqref{groupop} does not preserve geodesics, as discussed in~detail by the authors of~\cite{1}~-- the difference leads them to pose the integrability question also for the two-body case. In what follows, we give a decisive answer: the two-body problem on the Heisenberg group is not integrable.

For two point masses $m_1$ and $m_2$, whose positions are group
elements $g_k = (x_k,y_k,z_k)$ we will take the Hamiltonian to be
\begin{gather}
 H = \frac{1}{2m_1}\bigg(\bigg(p_{x_1}-\frac12 y_1 p_{z_1}\bigg)^2
 + \bigg(p_{y_1}+\frac12 x_1 p_{z_1}\bigg)^2\bigg)\nonumber
 \\ \hphantom{H =}
 {}+ \frac{1}{2m_2}\bigg(\bigg(p_{x_2}-\frac12 y_2 p_{z_2}\bigg)^2 +
 \bigg(p_{y_2}+\frac12 x_2 p_{z_2}\bigg)^2\bigg) +
 V\big(\rho\big(g_1^{-1}\cdot g_2\big)\big),\label{eq:2b}
\end{gather}
where the potential is specified by
\begin{equation*}
 V(\rho) = -\frac{\kappa m_1 m_2}{\rho},\qquad
 \rho(g) = \sqrt{\big(x^2+y^2\big)^2+{16}z^2}.
\end{equation*}

We note that the following first integrals are ``known'':
\begin{alignat*}{3}
& I_1 = p_{x_1} +\frac12 y_1 p_{z_1} + p_{x_2} +\frac12 y_2 p_{z_2},
 \qquad&&
 I_2 = p_{y_1} -\frac12 x_1 p_{z_1} + p_{y_2} -\frac12 x_2 p_{z_2},&
 \\
 &I_3 = p_{z_1}+p_{z_2}=\{I_1,I_2\},
 \qquad&&
 I_4 = y_1 p_{x_1}-x_1 p_{y_1} + y_2 p_{x_2} - x_2 p_{y_2},&
\end{alignat*}
and they satisfy
\begin{equation*}
 \{I_1,I_4\}=I_2, \qquad
 \{I_2,I_4\} = -I_1.
\end{equation*}
Additionally, as for the Kepler--Heisenberg problem,
\begin{equation*}
 J = x_1 p_{x_1} + y_1 p_{y_1} + 2 z_1 p_{z_1} + x_2 p_{x_2} + y_2 p_{y_2} + 2 z_2 p_{z_2},
\end{equation*}
is such that $\dot{J}=2H$. That is, we have 5 first integrals,
although they do not all commute, and on the zero-energy level $J$
becomes the sixth integral. The question, as before, is whether there
exist enough (here: six) commuting integrals.

Now, we make linear canonical transformation
\begin{alignat*}{3}
 &u_1 = \frac{1}{\sqrt{2}}(y_1-\mathrm{i} x_1),\qquad&&
 p_{u_1} = \frac{1}{\sqrt{2}}(p_{y_1}+\mathrm{i}p_{x_1}),&
 \\
& v_1 = \frac{1}{\sqrt{2}}(y_1+\mathrm{i} x_1),\qquad&&
 p_{v_1} =\frac{1}{\sqrt{2}}(p_{y_1}-\mathrm{i}p_{x_1}),&
 \\
& w_1 = z_1 +z_2, \qquad &&
 p_{w_1} = \frac12(p_{z_1}+p_{z_2}),&
 \\
 &u_2 = \frac{1}{\sqrt{2}}(y_2-\mathrm{i} x_2),\qquad&&
 p_{u_2} = \frac{1}{\sqrt{2}}(p_{y_2}+\mathrm{i}p_{x_2}),&
 \\
 &v_2 = \frac{1}{\sqrt{2}}(y_2+\mathrm{i} x_2),\qquad&&
 p_{v_2} =\frac{1}{\sqrt{2}}(p_{y_2}-\mathrm{i}p_{x_2}),&
 \\
& w_2 = z_1 -z_2, \qquad&&
 p_{w_2} = \frac12(p_{z_1}-p_{z_2}).&
\end{alignat*}
In the new variables the Hamiltonian reads
\begin{gather*}
 H = \frac{((p_{w_1}+p_{w_2})u_1-2\mathrm{i}p_{v_1})
 ((p_{w_1}+p_{w_2})v_1+2\mathrm{i}p_{u_1})}{4m_1}
 \\ \hphantom{ H =}
{} + \frac{((p_{w_1}-p_{w_2})u_2-2\mathrm{i}p_{v_2})
 ((p_{w_1}-p_{w_2})v_2+2\mathrm{i}p_{u_2})}{4m_2} -\frac{\kappa m_1 m_2}{\rho_{12}},
\end{gather*}
where
\begin{equation*}
 \rho_{12} = 2\sqrt{(u_1-u_2)^2(v_1-v_2)^2-(v_1 u_2 - v_2 u_1 + 2\mathrm{i}w_2)^2}.
\end{equation*}

The particular solution is almost as before
\begin{equation*}
 p_{w_1}=p_{w_1}(0),\qquad
 p_{w_2} = -2at,\qquad
 a = \frac{m_1 m_2\kappa}{8w_2|w_2|},\qquad
 w_2=w_2(0), \qquad
 w_1=w_1(0),
\end{equation*}
and all other phase variables equal to zero. The solution must not be constant, so $w_2(0)\neq 0$, but other parameters are not restricted.

Linear variations of the variables
$(u_1, p_{v_1},u_2,p_{v_2},v_1,p_{u_1},v_2,p_{u_2},w_1, w_2,
p_{w_1},p_{w_2})$, which we will denote by $\xi=(\xi_1, \ldots, \xi_{12})$, then
satisfy the variational equations
\begin{equation}
 \label{eq:varg}
 \dot \xi = A\xi, \qquad A = \begin{bmatrix}
 A_1 & 0 & 0\\
 0 & A_2 & 0\\
 0 & 0 & A_3
 \end{bmatrix}\!,
\end{equation}
where
\begin{equation*}
 A_1 = \begin{bmatrix}
 \tau -\tau_0& 1 & 0 & 0\\
 (\tau-\tau_0)^2 & \tau-\tau_0 & -1 & 0\\
 0 & 0 & -\mu(\tau+\tau_0) & \mu \\
 1 & 0 & \mu(\tau+\tau_0)^2 & -\mu(\tau+\tau_0)
 \end{bmatrix}\!,\qquad
 A_3 = \begin{bmatrix}
 0 & 0 & 0 & 0\\
 0 & 0 & 0 & 0\\
 0 & 0 & 0 & 0 \\
 0 & 4\mathrm{i}/w_2 & 0 & 0
 \end{bmatrix}\!,
\end{equation*}
and
\begin{equation*}
 A_2 = \begin{bmatrix}
 \tau_0-\tau & 1 & 0 & 0\\
 (\tau_0-\tau)^2 & \tau_0-\tau & 1 & 0\\
 0 & 0 & \mu(\tau+\tau_0) & \mu \\
 -1 & 0 & \mu(\tau+\tau_0)^2 & \mu(\tau+\tau_0)
 \end{bmatrix}\!.
\end{equation*}
To obtain the above form we use the following rescalings
\begin{equation*}
 t = m_1\tau,\qquad
 a = \frac{\mathrm{i}}{m_1},\qquad
 \mu = \frac{m_1}{m_2}, \qquad
 p_{w_1}=2 \mathrm{i} \tau_0.
\end{equation*}

\begin{Theorem} %\label{thm:1}
 If $\mu\neq -1$ then the two-body problem on the Heisenberg group is
 not integrable in the Liouville sense.
\end{Theorem}
\begin{proof}
 If the system generated by~\eqref{eq:2b} is integrable then by
 Theorem~\ref{thm:MoRa}, the identity component of differential
 Galois group of variational equations~\eqref{eq:varg} is Abelian.
 This implies that the same property is shared by the differential Galois groups
 of the subsystems of~\eqref{eq:varg}, which have the form
 $\dot\eta=A_i \eta$, $\eta\in\mathbb{C}^4$, for $i=1,2,3$. We
 consider the first of them. It has particular solution
 \begin{equation}
 \label{par}
 \eta(\tau)=(1, \tau_0-\tau,1, \tau_0+\tau).
 \end{equation}
 Using the d'Alambert method, see~\cite{Walter}, we can reduce the
 dimension of the system by one. But~assuming that $\tau_0=0$ we
 achieve more. Namely, linear transformation $\eta\mapsto Q\eta$ with
 $Q=Q(\tau)$ given by
 \begin{equation*}
 Q = \begin{bmatrix}
 1 & 0 & 0 & 0\\
 -\tau & 1 & 0 & 0\\
 1 & 2\tau & -1 & 0\\
 \tau & -1-2\tau^2 & \tau & -\frac{1}{\mu}
 \end{bmatrix}\!,
 \end{equation*}
 brings it to the form $\dot{\eta}=\widetilde{A}_1\eta$, where
 \begin{equation*}
 \widetilde{A}_1 =Q^{-1}\bigg(A_1 Q -\frac{\mathrm{d}}{\mathrm{d}\tau}Q\bigg) = \begin{bmatrix}
 0 & 1 & 0 & 0\\
 0 & 0 & 1 & 0\\
 0 & p_2 & p_1 & 1\\
 0 & 0 & 0 & 0
 \end{bmatrix}\!,\qquad p_1 = 2(1-\mu)\tau,\qquad p_2 =
 3+\mu+4\mu\tau^2.
 \end{equation*}
 Thus the transformed system has block-triangular structure which is
 quite simple: the first coordinate does not enter, while the fourth
 is constant. In other words, to obtain a particular solution, it is
 enough to assume $\eta_4=0$, and choose the subsystem corresponding to the second and third
 components:
 \begin{gather*}
 \eta_2'(\tau) = \eta_3(\tau),
 \\
 \eta_3'(\tau) = p_2(\tau) \eta_2(\tau) + p_1(\tau)\eta_3(\tau).
 \end{gather*}
 As a single equation it reads
 \begin{equation*}
 \eta_2'' = 2(1-\mu)\tau \eta_2' + \big(3+\mu+4\mu \tau^2\big)\eta_2,
 \end{equation*}
 which, after the change $\eta_2=\exp\big[(1-\mu)\tau^2/2\big]w(\tau)$, becomes
 \begin{equation}
 \label{eq:w2b}
 w''(\tau) -(1+\mu)\big[2 +(1+\mu)\tau^2\big]w(\tau)=0.
 \end{equation}
 It is, again, the parabolic cylinder equation~\eqref{eq:5} with parameters
 \begin{equation*}
 \alpha^2= (1+\mu)^2, \qquad
 \beta=0, \qquad
 \gamma = 2(1+\mu).
 \end{equation*}
 Let us assume that $\mu\neq-1$. Then $\alpha\neq 0$ and
 \[
 \frac{\beta^2-\gamma}{\alpha}=-2\sgn(1+\mu)
 \]
 is not an odd integer. Hence, by Theorem~\ref{thm:par} the
 differential Galois group of equation~\eqref{eq:w2b} is~$\mathrm{SL}(2,\mathbb{C})$. This ends the proof.
\end{proof}

The case $\mu=-1$ is difficult to study. Considering variational
equations with $\tau_0=0$ we do not obtain any obstacles for
integrability. Moreover, taking non-zero $\tau_0$ we are unable to
reduce the problem to study a second order differential equation.
Nevertheless we are able to show the following.

\begin{Theorem}%* \label{thm:2}
 If $\mu=-1$ then the two-body problem on the Heisenberg group is not
 integrable in the Liouville sense.
\end{Theorem}
\begin{proof}
 As in the previous proof we consider subsystem of variational
 equations~\eqref{eq:varg} corresponding to the matrix $A_1$ but now
 we fix $\mu=-1$ and $\tau_0=1$. Then, using particular
 solution~\eqref{par} we reduce its dimension to 3. But now to achieve
 this we make linear transformation $\eta\mapsto Q\widetilde\eta$
 with $Q=Q(\tau)$ given by
 \begin{equation*}
 Q = \begin{bmatrix}
 1 & 0 & 0 & 0\\
 \tau-1 & 1 & 0 & 0\\
 -1 & 0 & -1 & 0\\
 \tau +1 & -1 & \tau+1 & 1
 \end{bmatrix}\!.
 \end{equation*}
 We get $\dot{\eta}=\widetilde{A}_1\eta$, where
 \begin{equation*}
 \widetilde{A}_1 =Q^{-1}\bigg(A_1 Q -
 \frac{\mathrm{d}}{\mathrm{d}\tau}Q\bigg) =
 \begin{bmatrix}
 2(\tau -1) & 1 & 0 & 0\\
 0 & 0 & 1 & 0\\
 4 & -2 & 2(\tau+1) & 1\\
 0 & 0 & 0 & 0
 \end{bmatrix}\!.
 \end{equation*}
 Assuming that $\eta_4=0$ we consider system of first three equations
 \begin{equation}
 \label{s3}
 \begin{bmatrix}
 \eta_1'\\
 \eta_2'\\
 \eta_3'
 \end{bmatrix}=\begin{bmatrix}
 2(\tau -1) & 1 & 0 \\
 0 & 0 & 1 \\
 4 & -2 & 2(\tau+1)
 \end{bmatrix} \begin{bmatrix}
 \eta_1\\
 \eta_2\\
 \eta_3
 \end{bmatrix}\!.
 \end{equation}
 It is important to notice that the only singularity of this system
 is $\tau=\infty$, so all its solutions are holomorphic on the whole
 complex plane. We prove that it does not have any Liouvillian
 solution and thus its differential Galois group is not solvable. To
 apply conditions formulated in \cite{SU}, we rewrite
 system~\eqref{s3} as the third order equation
 \begin{equation*}
% \label{o3}
 \eta_2''' -4\tau\eta_2''+ 4\big(\tau^2-1\big)\eta_2' -4\tau \eta_2 = 0,
 \end{equation*}
 and then we substitute $\eta_2 = v(\tau) \exp\big[2\tau^2/3\big]$. As the
 result we obtain equation
 \begin{equation}
 \label{o3r}
 v''' -\frac{4}{3}\tau^2v'+ \frac{4}{27}\tau\big(4\tau^2-63\big)v = 0,
 \end{equation}
 whose differential Galois group is a subgroup of
 $\mathrm{SL}(3,\mathbb{C})$. According to \cite{SU} if this equation
 has a Liouvillian solution, then there are three possibilities:
 \begin{enumerate}\itemsep=0pt
 \item[$(1)$] it has a solution whose logarithmic derivative $v'/v$ is
 rational, or
 \item[$(2)$] it has three linearly independent solutions whose logarithmic
 derivative $v'/v$ are algebraic of order 3, or
 \item[$(3)$] all its solutions are algebraic.
 \end{enumerate}
 If none of the above cases occur, then the equation has no
 Liouvillian solution. Unfortunately, a direct application of the
 ``necessary conditions for case 1'' given in~\cite[p.~9]{SU} shows
 that these conditions are fulfilled. In order to exclude this case
 we have to use the full algorithm for checking if the equation admits an
 exponential solution, or just use a computer algebra system to check
 it. We use the Maple algebra system function \verb+exp_sol+ applied
 to equation~\eqref{o3r}, and it does not give any exponential
 solution.

 The equation is not Fuchsian~-- with one irregular singular point at
 infinity. This is why the third case is excluded.

 According to ``necessary conditions for case 2'' given in
 \cite[p.~12]{SU}, if this case occurs then the third symmetric power of
 equation~\eqref{o3r} has a solution of the form
 \begin{equation}
 \label{eq:11}
 v=P(\tau)\prod_{i=1}^{s}(\tau-\tau_i)^{\alpha_i},
 \end{equation}
 where $P(\tau)$ is a polynomial, $\tau_i$ is a singular point, and
 $\alpha_i$ is an exponent at this point. Moreover~$\alpha_i$ is a
 half integer for $i=1,\ldots, s$. Calculations, with the help of
 Maple, show that the third symmetric power of equation~\eqref{o3r} is
 an equation of order $10$ which has 15 regular singular points
 $\tau_i\in\mathbb{C}$. They are roots of the following polynomial
 \begin{gather*}
 S(\tau):=\tau\big(3456\,{\tau}^{14}-271680\,{\tau}^{12}+ 8200960\,{\tau}^{10}-119918560\, {\tau}^{8}
 +854800080\,{\tau}^{6}
 \\ \hphantom{ S(\tau):=\tau\big(}
 {}-2391850656\,{\tau}^{4}+ 71751150\,{ \tau}^{2}-229734225\big).
 \end{gather*}
 At each of these points $\alpha_i\in\{0,1,2,3,4,5,6,7,8,10\}$. The
 infinity is an irregular singular point with only one exponent
 $\alpha_{\infty}=2$. From the above facts it follows that if a
 solution of the form~\eqref{eq:11} exists then it is a polynomial,
 but then there must be an exponent at infinity which is equal to
 minus the degree of this polynomial. As there is no such exponent
 the second case does not occur. To conclude, equation~\eqref{o3r}
 does not admit any Liouvillian solution, so the identity component
 of its differential Galois group is not Abelian.
\end{proof}

\section{Concluding remarks}

Our main goal was to answer the question of Montgomery and Shanbrom about integrability of~the (simple) Kepler problem on the Heisenberg group. The answer turned out to be negative, but several generalisations became immediately apparent. First, the potential had a specific radial/axial symmetry, and a whole general class of such potentials could be included; second, and more important, the two-body problem could be formulated in a natural way. We thus extended the analysis, and managed to show, that with reasonable assumptions those extensions were also non-integrable.

We note that potentials not satisfying condition \eqref{eq:con} can be found, such as
\begin{equation*}
 V = z^2\big(x^2+y^2\big)^2 = z^2\big(\rho^2-16z^2\big), \qquad\text{or}\qquad V=\big(x^2+y^2\big)^2R(z,\rho),
\end{equation*}
where $R(z,\rho)$ is not divisible by $\big(x^2+y^2\big)^2$. Integrability of these potentials remains an open question.
One possible way of investigation of such cases is the application of a variant of the direct method. However, we were unable to find any integrals which were polynomials of low degree in momenta. It remains an open question whether our result can be extended to a wider functional class of first integrals, but each case requires a completely different set of methods than those used here, and as such is a subject for separate investigation.

\appendix

\section{Systems with algebraic Hamiltonians}\label{App0}

First, let us remark that Theorem~\ref{thm:MoRa} also holds for a
general Poisson system \cite{akn}. When the Hamiltonian function is algebraic but not meromorphic,
we cannot apply this theorem directly. One solution is to find an extension of the phase space (by including additional variables) in~such a way, that the original Hamiltonian lifts to a meromorphic one, and the extended system is Hamiltonian
with respect to a degenerate Poisson bracket, which reproduces
the original problem.

The construction below is a modification of that given in
\cite{alg}, where the reader will find more details and proofs. Let us
consider an $n$ degrees of freedom Hamiltonian system with canonical
coordinates $q,p\in\mathbb{C}^n$, with algebraic Hamiltonian $H(q,p)$
such that $H(q,p)=K(q,p,u)$, where~$u$ is algebraic over
$\mathbb{C}(q)$ with minimal polynomial $P(u)\in\mathbb{C}(q)[u]$, and
$K(q,p,u)\in \mathbb{C}(q,p,u)$ is a~rational function of its
arguments $x=(q,p,u)\in\mathbb{C}^{2n+1}$. We introduce the following
system
\begin{equation}
 \label{eq:dK}
 \dot x = J(x) \nabla_xK(x),
\end{equation}
where $J(x)$ is $(2n+1)\times(2n+1)$ matrix of the form
\begin{equation*}
 J(x):=\begin{bmatrix}
 0 & \mathbbm{1}_n & 0 \\
 -\mathbbm{1}_n & 0 & \frac{1}{\partial_u P}\nabla_q P \\
 0 & -\frac{1}{\partial_u P} \nabla_q P& 0
 \end{bmatrix}\!,
\end{equation*}
with $\mathbbm{1}_n$ equal to the $n\times n$ identity matrix. It defines the Poisson bracket
\begin{equation*}
 \{f,g\}(x) := (\nabla_xf(x))^{\rm T} J(x)\nabla_xg(x),
\end{equation*}
where $f$ and $g$ are smooth functions. The rank of matrix $J(x)$ is
$2n$ and the only Casimir function of the bracket is $P(u)$.
\begin{Lemma}
{\sloppy If $(q(t),p(t),u(t))$ is a solution of equations~\eqref{eq:dK} with
 $P(u(t))=0$, then $(q(t),p(t))$ is a solution of Hamilton's
 equations
 \begin{equation*}
 %\label{eq:has}
 \dot q= \nabla_q H(q,p), \qquad
 \dot p=-\nabla_p H(q,p).
 \end{equation*}}
\end{Lemma}
We omit the proof, as it is rather direct, and ask the interested reader to follow the explanation and steps given in \cite[Section~2]{alg}. This lemma gives us what is needed, that is we reproduce the original system as a~Hamiltonian one with respect to a degenerate Poisson structure defined
by rational matrix $J(x)$, and with rational Hamiltonian function~$K(x)$.

The above general considerations justify the meromorphic assumptions
of the Morales--Ramis theory, but of course for practical purposes the
calculations can be performed in the original coordinates.

\section{The factorization algorithm}\label{App1}

We outline the reduction of the variational system \eqref{VE}, following the notation of \cite{Weil}.

Take the nontrivial block of the VE, with $a=2$ (specific $z_0$ in the
particular solution), which~is
\begin{equation}
 \dot{\eta} =
 \begin{bmatrix} 0 & 1 & 2t & 0\\
 -4t^2 & 0 & 0 & 2t \\
 -2t & 0 & 0 & 1 \\
 0 & -2t & -4t^2 & 0 \end{bmatrix}\eta,
 \label{Weil_VE}
\end{equation}
and construct the associated system, which is
its second external power, i.e., the differential equation
for an antisymmetric matrix $W$, which reads
\begin{equation*}
 \dot{W} = A W - W^{\rm T} A^{\rm T},
\end{equation*}
where $A$ is the coefficient matrix in \eqref{Weil_VE}. The matrix $W$
has 6 components, so we are effectively dealing with a six-dimensional
linear system
\begin{equation*}
\dot{Y} =
 \begin{bmatrix}
 0 & 0 & 2t & -2t & 0 & 0 \\
 0 & 0 & 1 & 1 & 0 & 0\\
 -2t & -4t^2 & 0 & 0 & 1 & 2t \\
 2t & -4t^2 & 0 & 0 & 1 & 2t \\
 0 & 0 & -4t^2 & -4t^2 & 0 & 0 \\
 0 & 0 & -2t & 2t & 0 & 0
 \end{bmatrix}Y.
\end{equation*}
The next task in the algorithm is to find an exponential solution $Y$. In the above matrix, the third and fourth rows (and columns), can be combined to eliminate some of the $t^2$ terms, and a~simple basis permutation gives the similarity transform to the following block diagonal form
\begin{equation*}
 \begin{bmatrix}
 A_1 & 0 \\ 0 & A_2
 \end{bmatrix},\qquad
 A_1 = t \begin{bmatrix}
 0 & -2 & 0\\
 4 & 0 & -4\\
 0 & 2 & 0
 \end{bmatrix}\!,\qquad
 A_2 =
 \begin{bmatrix}
 0 & t^2 & 0\\
 -8 & 0 & t^2\\
 0 & -8 & 0
 \end{bmatrix}\!.
\end{equation*}
The first block can be solved with the exponential factor
$\exp\big({\pm}2\mathrm{i}t^2\big)$, and, surprisingly, we recover two solutions
of the associated system in one step. They read:
\begin{gather*}
 Y_1 = \exp\big(2\mathrm{i}t^2\big) [ -1, 0, -\mathrm{i}, \mathrm{i}, 0, 1]^{\rm T},
 \qquad
 Y_2 = \exp\big({-}2\mathrm{i}t^2\big) [ -1, 0, \mathrm{i}, -\mathrm{i}, 0, 1]^{\rm T}.
\end{gather*}

We next check the Pl\"ucker condition
$z_{03} z_{12}-z_{02}z_{13}+z_{23}z_{01}=0$, taking for each solution
$Y_k = [z_{01},z_{02},z_{03}, z_{12},z_{13}, z_{23}]$. In our case, it
is trivially satisfied for each $Y_k$, and that means that the respective operators
\begin{equation*}
 M_{\Psi} = \begin{bmatrix}
 z_{12} & - z_{02} & z_{01} & 0\\
 z_{13} & -z_{03} & 0 & z_{01}\\
 z_{23} & 0 & -z_{03} & z_{02}\\
 0 & z_{23} & -z_{13} & z_{12}
 \end{bmatrix}
\end{equation*}
have non-trivial kernels spanned by some $\{e_i\}$~-- these need to be combined, and possibly completed, to form the new basis.
Each kernel is two-dimensional here, so we get $e_1$ and $e_2$ from~$Y_1$,
and $e_3$ and $e_4$ from~$Y_2$, which can be collected as columns in
the full basis
\begin{equation*}
 Q = \begin{bmatrix}
 0 & -\mathrm{i} & 0 & \mathrm{i}\\
 -\mathrm{i} & 0 & \mathrm{i} & 0\\
 0 & 1 & 0 & 1\\
 1 & 0 & 1 & 0
 \end{bmatrix}\!, \qquad \det(Q)=-4.
\end{equation*}
Making the transformation, turns the VE into the block-diagonal form
\begin{equation}
\label{a1}
 Q^{-1}AQ = \begin{bmatrix}
 2\mathrm{i}t & -4t^2 & 0 & 0\\
 1 & 2\mathrm{i} t & 0 & 0\\
 0 & 0 & -2\mathrm{i} t & -4t^2 \\
 0 & 0 & 1 & -2\mathrm{i}t
 \end{bmatrix}\!.
\end{equation}
That the form is not merely block-triangular is thanks to the previous step yielding, by chance, enough of the $e_i$.

Note that $Q$ works regardless of the value of $a$, so it can immediately be lifted to a linear canonical transformation in the original variables:
\begin{alignat*}{3}
& u = \frac{1}{\sqrt{2}}(y-\mathrm{i} x),\qquad&&
 p_u = \frac{1}{\sqrt{2}}(p_y+\mathrm{i}p_x),&
 \\
& v = \frac{1}{\sqrt{2}}(y+\mathrm{i} x),\qquad&&
 p_v =\frac{1}{\sqrt{2}}(p_y-\mathrm{i}p_x),&
%\label{canon}
\end{alignat*}
after which the Hamiltonian becomes
\begin{equation*}
 H = \bigg(\frac12 u p_z -\mathrm{i}p_v\bigg)
 \bigg(\frac12 v p_z +\mathrm{i}p_u\bigg)
 -\frac{\alpha}{2\sqrt{u^2v^2+z^2}}.
\end{equation*}
The VE along our particular solution are block-diagonal in the variables $[u,p_v,v,p_u]$, but we note that they are quadratic in time, as in \eqref{a1}. This can be further simplified, by adding non-linear terms in the transformation of the original variables, as is done in the main text in~\eqref{eq:1} leading to linear VE in~\eqref{eq:3}.

\subsection*{Acknowledgments}

We would like to thank the anonymous referees for helping to improve the manuscript.
This work has been supported by grants No.\,DEC-2011/02/A/ST1/00208 and
DEC-2013/09/B/ST1/04130 of National Science Centre of Poland.
For the second author this research was partially supported by The National Science Center of Poland Under Grant No. 2020/39/D/ST1/01632.
%For the purpose of Open Access, the author has applied a CC-BY public copyright license to any Author Accepted Manuscript (AAM) version arising from this submission.

\pdfbookmark[1]{References}{ref}
\LastPageEnding

\end{document}